# Observation of inverse magnetocaloric effect in magnetic-field-induced austenite phase of Heusler Alloys $Ni_{50-x}Co_xMn_{31.5}Ga_{18.5}$ ($x$ = 9 and 9.7)

T. Kihara[1]*, T. Roy[2], X. Xu[3], A. Miyake[4], M. Tsujikawa[2], H. Mitamura[4] M. Tokunaga[4], Y. Adachi[5], T. Eto[6], and T. Kanomata[7]

[1] Institute for Materials Research, Tohoku University, Sendai, Miyagi, Japan
[2] Research Institute of Electrical Communication, Tohoku University, Sendai, Miyagi, Japan
[3] Department of Materials Science, Tohoku University, Sendai, Miyagi, Japan
[4] The Institute for Solid State Physics, The University of Tokyo, Kashiwa, Chiba, Japan
[5] Graduate School of Science and Engineering, Yamagata University, Yonezawa, Yamagata, Japan
[6] Kurume Institute of Technology, Kurume, Fukuoka, Japan
[7] Research Institute for Engineering and Technology, Tohoku Gakuin University, Tagajo, Miyagi, Japan

* Address all correspondence to: t_kihara@imr.tohoku.ac.jp

**ABSTRACT**

Magnetocaloric effect (MCE), magnetization, specific heat, and magnetostriction measurements were performed in both pulsed and steady high magnetic fields to investigate the magnetocaloric properties of Heusler alloys $Ni_{50-x}Co_xMn_{31.5}Ga_{18.5}$ ($x$ = 9 and 9.7). From direct MCE measurements for $Ni_{41}Co_9Mn_{31.5}Ga_{18.5}$ up to 56 T, a steep temperature drop was observed for magnetic-field-induced martensitic transformation (MFIMT), designated as inverse MCE. Remarkably, this inverse MCE is apparent not only with MFIMT, but also in the magnetic-field-induced austenite phase. Specific heat measurements under steady high magnetic fields revealed that the magnetic field variation of the electronic entropy plays a dominant role in the unconventional magnetocaloric properties of these materials. First-principles based calculations performed for $Ni_{41}Co_9Mn_{31.5}Ga_{18.5}$ and $Ni_{45}Co_5Mn_{36.7}In_{13.3}$ revealed that the magnetic-field-induced austenite phase of $Ni_{41}Co_9Mn_{31.5}Ga_{18.5}$ is more unstable than that of $Ni_{45}Co_5Mn_{36.7}In_{13.3}$ and that it is sensitive to slight tetragonal distortion. We conclude that the inverse MCE in the magnetic-field-induced austenite phase is realized by marked change in the electronic entropy through tetragonal distortion induced by the externally applied magnetic field.

**I. INTRODUCTION**

Magnetocaloric effect (MCE) is a spontaneous temperature change of magnetic materials through application or removal of an external magnetic field, which is a consequence of variation of entropy in the magnetic field. This effect has been believed to occur with a change of magnetic entropy. Application of a magnetic field to a paramagnetic or ferromagnetic material reduces the disorder of spins and thus the magnetic entropy. In the adiabatic magnetization process, the lattice entropy increases simultaneously to conserve the total entropy. Consequently, the material temperature increases (direct MCE). However, the magnetic field application along the magnetization easy axis of an antiferromagnetic or ferrimagnetic material destabilizes antiparallel spin sublattices and raises the magnetic entropy. In the adiabatic magnetization process, the material temperature decreases (inverse MCE) [1–6]. For inverse MCEs, the application of a sufficiently high magnetic field aligns the spins and reduces the magnetic entropy again. Consequently, inverse MCE is apparent in the low field region. It is usually smaller than the direct MCE [3,4].

Magnetic refrigeration (MR) based on the MCE developed independently by Debye [7] and Giauque [8] has long been used to realize extremely low temperatures. However, triggered by the discovery of the giant MCE [9], MR has attracted much attention as a key system to realize energy-efficient and environmentally safe refrigeration technologies functioning at around room temperature [10]. Giant MCE appears as a result of simultaneous entropy changes in the different degrees of freedom (spin, lattice, orbital, etc.). Consequently, the total entropy change can exceed the limit of the magnetic entropy change attributable to the forced spin alignment. Such simultaneous changes in different degrees of freedom are usually attained through the first-order phase transitions [9–18]. Therefore, investigators have devoted great attention to materials showing first-order phase transition with the purpose of exploiting entropy changes in different degrees of freedom.

For this purpose, Ni-Mn-based Heusler alloys Ni-Mn-$Z$ ($Z$ = Ga, In, Sn, and Sb) have been studied extensively during the last decade as a class of promising magnetocaloric materials [17–24]. These alloys have a cubic ($L2_1$) Heusler structure with a space group of $Fm\bar{3}m$ [18]. The non-stoichiometric compounds with low Z concentration undergo first-order martensitic transformation (MT) from a cubic austenite phase (A phase) to a martensite phase of a lower symmetry (M phase: tetragonal, orthorhombic, or monoclinic) with decreasing temperature. In these alloys, the structural and the magnetic properties strongly depend on the Z species and its compositional ratio. The ground state of the parent compound $Ni_{50}Mn_{50}$ is the antiferromagnetic tetragonal ($L1_0$) phase. The substitution of Mn with the third element Z progressively decreases the antiferromagnetic coupling between the spins and increases the lattice constant, which can stabilize the ferromagnetic cubic phase. Consequently, the MT temperature decreases significantly concomitantly with increasing Z concentration. For the stoichiometric $Ni_{50}Mn_{25}Z_{25}$ (Z = In, Sn, and Sb), the ferromagnetic A phase is the ground state [18]. $Ni_{50}Mn_{25}Ga_{25}$ is the only Ni-Mn-based Heusler alloy that shows the MT in the stoichiometric composition, which might be attributed to the relatively small ion radius of Ga [25]. The magnetic property of the M phase also varies with Z species. While the Ni-Mn-Z (Z = In, Sn, and Sb) compounds show the weakly magnetic M phase, Ni-Mn-Ga compounds show the ferromagnetic M phase with the larger saturation magnetization than that of the A phase [25]. The Z species dependence of the structural and the magnetic properties can comprehensively be understood through the valence electron concentration per atom ($e/a$) [18].

Partial substitution of Ni with Co enhances the ferromagnetic coupling between the spins, which increases the Curie temperature and decreases the MT temperature. Therefore, the quaternary compounds can undergo MT accompanied by the metamagnetic transition. For instance, $Ni_{45}Co_5Mn_{36.7}In_{13.3}$ undergoes MT from the ferromagnetic A phase to the paramagnetic M phase [19]. When a magnetic field is applied to the M phase, this alloy undergoes the magnetic-field-induced martensitic transformation (MFIMT) accompanied with the great increase in entropy, which is called inverse MCE [19,20]. The inverse MCEs are observed for the ternary and the Co-doped Ni-Mn-Z (Z = Ga, In, Sn, and Sb) compounds [17–24,26,27]. We earlier reported isothermal entropy change of 25 J/kg K through MT for $Ni_{45}Co_5Mn_{36.7}In_{13.3}$ [23]. The intensity of this inverse MCE is comparable to that of giant direct MCEs observed in Refs. [9,10,12–16].

Because the MTs in the Heusler alloys are realized through the band Jahn-Teller effect [28–30], they are accompanied by changes in the electron density of states (DOS), crystal structure, and magnetic structure. Therefore, electronic ($\Delta S_{ele}$), structural ($\Delta S_{lat}$), and magnetic ($\Delta S_{mag}$) entropy potentially contribute to the MCE, contrary to conventional MCEs in paramagnetic, ferromagnetic, antiferromagnetic, and ferrimagnetic materials. In such cases, the total entropy change ($\Delta S_{tot}$) is described as $\Delta S_{tot} = \Delta S_{ele} + \Delta S_{mag} + \Delta S_{lat}$. Consequently, inverse MCE is obtainable for $\Delta S_{ele} + \Delta S_{lat} > -\Delta S_{mag}$, even when the applied magnetic field aligns spins parallel to the field direction and reduces the magnetic entropy. For $Ni_{45}Co_5Mn_{36.7}In_{13.3}$, a comprehensive study of thermodynamic quantities using high magnetic fields revealed the great positive $\Delta S_{lat}$ at MT, which mainly causes the inverse MCE associated with MFIMT [23]. However, the electronic contribution to inverse MCE of $Ni_{45}Co_5Mn_{36.7}In_{13.3}$ is negligibly small, although a marked change in DOS at MT is expected intuitively by the band Jahn-Teller mechanism.

Here, we consider the possibility of realizing an inverse MCE driven by the large $\Delta S_{ele}$. According to the band Jahn-Teller mechanism, the MTs of the Ni-Mn-based Heusler alloys occur accompanied by the splitting of a DOS peak existing near the Fermi level in the cubic A phase. The drastic change in DOS through MT is expected theoretically for Ni-Mn-In alloys [30]. However, such a drastic change in DOS is not experimentally observed in the Co-doped $Ni_{45}Co_5Mn_{36.7}In_{13.3}$ [23]. As described in Sec. III E, this discrepancy between the theory and the experiment is mainly caused by the paramagnetic nature of the M phase in $Ni_{45}Co_5Mn_{36.7}In_{13.3}$. To enhance the electronic contribution to the MCE, we here focus on the Ni-Mn-Ga alloys. Because the lattice constant of the stoichiometric $Ni_{50}Mn_{25}Ga_{25}$ (~ 5.823 Å) is about 4% smaller than that of $Ni_{50}Mn_{25}In_{25}$ (~ 6.070 Å), the $d$ electrons in $Ni_{50}Mn_{25}Ga_{25}$ is more itinerant than that in $Ni_{50}Mn_{25}In_{25}$ [31,32]. Therefore, the comparison of the magnetocaloric properties between homologous Z = Ga and In enables quantitative evaluation of the electronic contribution to the MCE without significant change in the $e/a$.

A drastic change in DOS is expected theoretically for $Ni_{50}Mn_{25}Ga_{25}$ alloy [28,33–35]. In the ferromagnetic A phase of this alloy, a sharp peak of DOS exists near the Fermi level. When $Ni_{50}Mn_{25}Ga_{25}$ undergoes MT, this peak of DOS splits because of tetragonal distortion. Consequently, the DOS at the Fermi level decreases. Therefore, through the MFIMT from M to A phase, a great positive $\Delta S_{ele}$ attributable to the change in DOS is expected. The peak of DOS is also expected to exist in the A phase of the Co-doped Ni-Mn-Ga alloys [36,37].

This study specifically examines the Co-doped Heusler alloys $Ni_{50-x}Co_xMn_{31.5}Ga_{18.5}$ ($x$ = 9 and 9.7) to explore inverse MCE driven by the electronic entropy change. These alloys undergo MT from a ferromagnetic cubic A phase to a paramagnetic tetragonal M phase with decreasing temperature, as is the case for $Ni_{45}Co_5Mn_{36.7}In_{13.3}$. The MFIMTs were reported for both compounds [38]. We conducted a comprehensive study of magnetocaloric properties of these compounds by combining MCE, magnetization, specific heat, and magnetostriction measurements in both the pulsed and the steady high magnetic fields. Inverse MCEs at MFIMTs are measured directly for $Ni_{41}Co_9Mn_{31.5}Ga_{18.5}$. The inverse MCEs are also observed in the magnetic-field-induced A phase. Theoretical analysis using the first-principles calculations revealed that a great change in DOS contributes to the inverse MCE of the magnetic-field-induced A phase of $Ni_{41}Co_9Mn_{31.5}Ga_{18.5}$.

This paper is organized as follows. Section II briefly presents details of experiments and theoretical calculations. Section III provides experimentally and theoretically obtained results of magnetization (Sec. III A), MCE (Sec. III B), specific heat (Sec. III C), magnetostriction (Sec. III D), and DOS (Sec. III E). Section IV describes exploration of the microscopic origin of the inverse MCE of the magnetic-field-induced A phase. Section V presents conclusions inferred from our findings.

**II. EXPERIMENTS**

Polycrystalline samples of $Ni_{50-x}Co_xMn_{31.5}Ga_{18.5}$ ($x$ = 9 and 9.7) were synthesized using arc melting. Ingots that were vacuum-encapsulated in a quartz tube were annealed at 800 °C for 72 h. Subsequently they were quenched in cold water.

Using the nondestructive pulse magnet installed in the Institute for Solid State Physics of the University of Tokyo, we performed MCE and magnetostriction measurements for $Ni_{41}Co_9Mn_{31.5}Ga_{18.5}$. For the MCE measurements, a plate-shaped sample with thickness of less than 0.1 mm was used to reduce eddy current self-heating during the pulsed magnetic field application [39]. Instantaneous change of the sample temperature in the pulsed magnetic field was detected using a gold thin-film-resistive thermometer with 100 nm thickness, which is thermally contacted to the sample through the sapphire insulating layer. Further detailed descriptions of the MCE measurement are provided in earlier reports [39,40]. Magnetostriction measurements were conducted using the capacitance method [41,42]. Longitudinal magnetostrictions ($\mu_0 H \parallel \Delta L$) are measured using a cubic polycrystalline sample: $1.0 \times 1.0 \times 1.0$ mm$^3$.

Magnetization and specific heat measurements in steady magnetic fields were taken using superconducting magnets installed at the Institute for Materials Research of Tohoku University. Temperature variations of magnetization at constant magnetic fields up to 19 T were taken for $Ni_{41}Co_9Mn_{31.5}Ga_{18.5}$ using a vibrating sample magnetometer and a Quantum Design MPMS. Specific heat measurements of $Ni_{40.3}Co_{9.7}Mn_{31.5}Ga_{18.5}$ were conducted in magnetic fields up to 16 T using a quasi-adiabatic technique with a homemade probe [43]. Specific heat of around MT temperature at 0 T were also measured for $Ni_{41}Co_9Mn_{31.5}Ga_{18.5}$ and $Ni_{40.3}Co_{9.7}Mn_{31.5}Ga_{18.5}$ using a differential scanning calorimeter (DSC, 204F1 Phoenix; Netzsch Inc.).

Electronic, structural, and magnetic properties of $Ni_{50-x}Co_xMn_{31.5}Ga_{18.5}$ ($x$ = 9 and 9.7) have been probed using first-principles-based calculations. To obtain equilibrium lattice parameters, we calculated the total energy as a function of volume of their respective cubic unit cells using the Vienna *ab initio* Simulation Package (VASP) in combination with the projector augmented wave (PAW) method [44,45]. For each of the two systems, we constructed 100-atom special quasi-random structure (SQS) [46] using the alloy theory automated toolkit (ATAT) [47]. We used an energy cutoff of 500 eV for the plane waves, and the energies have been calculated with a $k$-mesh of 4 × 4 × 4 dimension. Generalized gradient approximation (GGA) has been used for the exchange correlation functional [48].

After obtaining the equilibrium geometry of the system using VASP, we obtained the electronic structure and the magnetic properties using Green function based full potential spin-polarized relativistic Korringa-Kohn-Rostoker method as implemented in the SPR-KKR program package [49,50]. The atomic disorder effect was considered in the coherent potential approximation (CPA). Here, we also used GGA for the exchange correlation functional. In all, 917 irreducible $k$-points have been used for the Brillouin zone integration and further an angular expansion up to $l_{max}$ = 3 has been used for each atom.

## III. RESULTS

### A. Magnetization

To determine the magnetic phase diagram and to evaluate magnetic entropy, we first present the $M$-$T$ curves of $Ni_{41}Co_9Mn_{31.5}Ga_{18.5}$ measured for several magnetic fields below 19 T, as shown in Fig. 1(a). For each magnetic field, the data are taken as follows: 300 K → 550 K → 300 K. A sharp magnetization jump corresponding to the thermal MT temperatures [$T_{M \to A}$ (heating process) and $T_{A \to M}$ (cooling process)] is apparent at around room temperature. The drop in magnetization with increasing temperature detected at around 470 K is attributable to the ferromagnetic-paramagnetic phase transition in A phase. Figure 1(b) portrays the temperature differential of the magnetization at 0.5 T. $T_{M \to A}$ = 352 K and $T_{A \to M}$ = 320 K are estimated respectively by the peaks of $dM/dT$. In addition, the Curie temperature of A phase ($T_C^A$) is estimated by the dip of $dM/dT$ as $T_C^A$ = 468 K. Similarly, the Curie temperature of M phase ($T_C^M$) is estimated as $T_C^M$ = 263 K from the $M$-$T$ curve below 300 K [Fig. 1(c)].

The magnetic phase diagram produced by the $M$-$H$ curves for $Ni_{41}Co_9Mn_{31.5}Ga_{18.5}$ is presented by open triangles in Fig. 2(a), where $H_{M \to A}$ and $H_{A \to M}$ are determined by the peaks of $dM/\mu_0 dH$ [38]. $T_{M \to A}$ and $T_{A \to M}$ shown by solid squares in Fig. 2(a) are consistent with the magnetic phase diagram produced by the $M$-$H$ curves. From this magnetic phase diagram, $\Delta S_{tot}$ at the MFIMT can be estimated using the Clausius-Clapeyron relation: $\mu_0 dH_0/dT = -\Delta S_{tot}/\Delta M$. Here, $H_0 = (H_{M \to A} + H_{A \to M})/2$ is assumed as the thermal equilibrium magnetic field. $\Delta M$ is the magnetization jump at the MFIMT. Results are portrayed in Fig. 2(b) as a function of the transformation temperature. It is noteworthy that the $\Delta S_{tot}$ of $Ni_{41}Co_9Mn_{31.5}Ga_{18.5}$ is smaller than that of $Ni_{45}Co_5Mn_{36.7}In_{13.3}$, which is consistent with the previous reports [51,52].

B. Magnetocaloric effect

We next present the results of the MCE measurements for $Ni_{41}Co_9Mn_{31.5}Ga_{18.5}$ in Figs. 3(a)-3(c). Here, the sample temperature is measured directly in an adiabatic condition. Because the sample is in M phase below 320 K, as described in Sec. III A, it undergoes MFIMTs at the transformation fields $H_{M \to A}$ (field increasing process) and $H_{A \to M}$ (field decreasing process). As presented in Fig. 3(a), clear temperature drops (i.e., inverse MCE) are observed at $T_{ini}$ = 238 - 301 K, as is true for $Ni_{45}Co_5Mn_{36.7}In_{13.3}$ [23], where $T_{ini}$ is the initial temperature at 0 T before the magnetic field application. Enlarged views of MCE are portrayed for the representative $T_{ini}$ values in Figs. 3(b) and 3(c), where $T_{ini}$ is subtracted from each curve [i.e., $\Delta T_{ad}(H) = T(H) - T_{ini}(0)$]. In the magnetic field increasing process, the sample temperature drops steeply at the fields indicated by downward-pointing arrows in Fig. 3(b); then, it is saturated above about 40 T. In successively measured magnetic field decreasing processes, the sample temperature increases reversibly at the beginning and then shows step-like behavior. For instance, when a magnetic field is applied at $T_{\text{ini}}$ = 301 K, the sample temperature decreases concomitantly with increase in strength of the magnetic field above 8 T; it saturates at around 44 T. The maximum $\Delta T_{\text{ad}}$ at 44 T reaches -14.9 K. Subsequently, with decreasing strength of the magnetic field from 50 T, the sample heats up almost reversibly at first. It shows a plateau in the field range of 2-8 T. Then, it heats up again below 2 T. Here, a small open loop is apparent above approximately 30 T in the data for $T_{\text{ini}}$ = 301 K, which results from the slight delay in response of the thermometer fabricated on the sample surface. Therefore, if the magnetic field with a slower sweep speed were applied, this small open loop would disappear. Because the field sweep speed of a pulsed magnetic field depends on the maximum field as shown in the inset of Fig. 3(a), when magnetic fields up to 37 T with about 34% slower sweep speed are applied at $T_{\text{ini}}$ = 263 and 280 K, the completely reversible MCE data are respectively obtained in the magnetic field range above 22 T and above 26 T, as presented in Fig. 3(c). It is noteworthy that such inverse MCEs above $H_{\text{M}\rightarrow\text{A}}$ (i.e., in the magnetic-field-induced A phase) are not observed for $Ni_{45}Co_5Mn_{36.7}In_{13.3}$. As shown by the gray dotted curve in Fig. 3(c), $Ni_{45}Co_5Mn_{36.7}In_{13.3}$ shows a positive slope in $\Delta T_{\text{ad}}$ in the magnetic field ranges above 12 T in the field increasing process and above 8 T in the field decreasing process. This direct MCE in the magnetic-field-induced A phase of $Ni_{45}Co_5Mn_{36.7}In_{13.3}$ is attributed to the negative change in the magnetic entropy caused by the forced spin alignment. The negative $\Delta T_{\text{ad}}$ values obtained for $Ni_{41}Co_9Mn_{31.5}Ga_{18.5}$ indicate that the entropy increases with increasing magnetic field, which cannot be explained by the forced spin alignment mechanism.

The magnetic phase diagram produced by the MCE data is shown by the solid triangles in Fig. 2(a), where $H_{\text{M}\rightarrow\text{A}}$ and $H_{\text{A}\rightarrow\text{M}}$ are defined respectively as the fields indicated by the downward-pointing arrows and upward-pointing arrows in Fig. 3(b). Results show good agreement with those produced by the magnetization data (open triangles), which also confirms the inverse MCE at MFIMT for $Ni_{41}Co_9Mn_{31.5}Ga_{18.5}$. The adiabatic temperature change is converted into the isothermal entropy change ($\Delta S_{tot}$) through the thermodynamic relation: $\Delta T_{ad} \cong -T\Delta S_{tot}/C$. Here, the experimentally obtained specific heat values ($C$) at 0 T shown in the inset of Fig. 4 are used to estimate $\Delta S_{tot}$ at MFIMT. Using $\Delta T_{ad}(44\ T) = -14.9$ K for $T_{ini} = 301$ K, we obtain $\Delta S_{tot} \sim 1.49$ J/mol K. Surprisingly, this value is significantly larger than that estimated from the magnetic phase diagram presented in Fig. 2(b). However, if one uses the temperature difference between $H_{M \to A}$ and $H_{A \to M}$, as indicated by the double-headed arrow in Fig. 3(c) [i.e., $\Delta T_{MT} = \Delta T_{ad}(H_{M \to A}) - \Delta T_{ad}(H_{A \to M})$] for the estimation of $\Delta S_{tot}$, then the results show good agreement with that estimated from the magnetic phase diagram, as shown by the open triangles in Fig. 2(b), which indicates that $\Delta T_{MT}$ corresponds to the MCE attributable to the MFIMT. In other words, the inverse MCEs appearing in the field range above $H_{A \to M}$ cannot be explained by the entropy change associated with the MFIMT derived from the Clausius-Clapeyron relation. Consequently, the inverse MCEs observed above $H_{A \to M}$ can be regarded as that in the magnetic-field-induced A phase.

Magnetic entropy change can be estimated by magnetization through the Maxwell relation: $\Delta S_{mag} = \int_0^H (\partial M/\partial T)_H dH'$. According to this relation, the sign of the magnetic entropy change depends on that of $\partial M/\partial T$. As depicted in Fig. 1(a), $\partial M/\partial T < 0$ in the A phase in the magnetic field range below 19 T. Evidently, the magnetic entropy of A phase ($S_{mag}^A$) decreases concomitantly with increasing magnetic field, which engenders a positive $\Delta T_{ad}$. Therefore, the negative $\Delta T_{ad}$ obtained above $H_{M \to A}$ cannot be explained solely by the magnetic entropy. As described in Sec. I, because the electronic, magnetic, and lattice entropy potentially contribute to the inverse MCE of $Ni_{41}Co_9Mn_{31.5}Ga_{18.5}$, we next evaluate electronic and lattice degrees of freedom.

**C. Specific heat**

To evaluate the electronic and lattice contributions to the total entropy change, specific heat measurements were performed in the steady high magnetic fields. Here, we used the extra-Co-doped alloy $Ni_{40.3}Co_{9.7}Mn_{31.5}Ga_{18.5}$ to lower the transformation field. As shown in the inset of Fig. 4, $Ni_{41}Co_9Mn_{31.5}Ga_{18.5}$ and $Ni_{40.3}Co_{9.7}Mn_{31.5}Ga_{18.5}$ show respectively the peaks at 363 K and 316 K attributable to MT. These peaks are remarkably smaller than that of $Ni_{45}Co_5Mn_{36.7}In_{13.3}$ and $Ni_{45}Co_5Mn_{36.5}In_{13.5}$, which is consistent with the small $\Delta S_{tot}$ estimated from the magnetic phase diagram. Because of the small peaks at MT, direct estimations of $\Delta S_{tot}$ from the specific heat data are difficult.

According to our earlier report, $Ni_{40.3}Co_{9.7}Mn_{31.5}Ga_{18.5}$ exhibits similar magnetic properties to those of $Ni_{41}Co_9Mn_{31.5}Ga_{18.5}$ [38]. Actually, $Ni_{40.3}Co_{9.7}Mn_{31.5}Ga_{18.5}$ remains in the A phase at low temperatures through field-cooling above 7 T [38]. Therefore, the specific heat in both A and M phases can be measured for the same sample. Figure 4 presents results of specific heat measurements plotted as $C/T$ vs. $T^2$. Data at 5, 10, and 16 T are taken after field-cooling at 17.5 T. Data at 0 T are taken after the zero-field-cooling. Here, $Ni_{0.403}Co_{0.097}Mn_{0.315}Ga_{0.185}$ is defined as the formula unit. The linear fit to the data for 0 and 16 T enables us to estimate, respectively, the Sommerfeld coefficient ($\gamma$) and the Debye temperature ($\Theta$) for both A (16 T) and M (0 T) phases in $Ni_{40.3}Co_{9.7}Mn_{31.5}Ga_{18.5}$: $\gamma_A$ = 5.3 mJ/mol K$^2$ (A phase), $\gamma_M$ = 3.2 mJ/mol K$^2$ (M phase), $\Theta_A$ = 331 K (A phase), and $\Theta_M$ = 362 K (M phase). It is noteworthy that both $\gamma$ and $\Theta$ show a remarkable difference between A and M phases, contrary to the case of $Ni_{45}Co_5Mn_{36.5}In_{13.5}$. For $Ni_{45}Co_5Mn_{36.5}In_{13.5}$, only the Debye temperature changes markedly at MT, which indicates that the lattice entropy change contributes dominantly to the inverse MCE in $Ni_{45}Co_5Mn_{36.5}In_{13.5}$ [23].

The difference of the electronic entropy between the A and M phase can be calculated immediately using $\Delta S_{ele}(T_{ini}) = (\gamma_A - \gamma_M)T_{ini}$. The obtained $\Delta S_{ele}$ at $T_{ini}$ = 300 K reaches 0.63 J/mol K, which exceeds the $\Delta S_{tot}$ presented in Fig. 2(b). However, the lattice entropy change can be estimated roughly using the Debye model. At $T_{ini}$ = 300 K, we obtain $\Delta S_{lat}$ = 2.1 J/mol K, which is considerably larger than the $\Delta S_{tot}$. Therefore, we conclude that both the electronic and lattice entropy play dominant roles in the inverse MCE of $Ni_{41}Co_9Mn_{31.5}Ga_{18.5}$.

**D. Magnetostriction**

According to the band Jahn-Teller model, both electronic and phonon band structures of the Ni-Mn-based Heusler alloys are sensitive to a symmetry-breaking structural distortion [28,29,31,35,53]. Therefore, we took the magnetostriction measurements of $Ni_{41}Co_9Mn_{31.5}Ga_{18.5}$ to investigate the magnetic field variation of the crystal structure. Figure 5 shows longitudinal magnetostrictions along the applied magnetic field ($\mu_0 H \parallel \Delta L$) measured at various temperatures. Here, the changes in the sample length ($\Delta L$) are taken in order of decreasing temperature. Positive $\Delta L$ is observed at the MFIMT as reported earlier by Sakon *et al*. [54,55]. The total strain decreases concomitantly with decreasing temperature, which indicates that the magnetic-field-induced A phase is partially arrested in M phase at 0 T. At 290 K, the total strain reaches 0.19%. Assuming isotropic magnetostriction for this polycrystalline sample, the total volume change can be estimated as 0.57%. Here, using the volume of the unit cell of M phase ($V_M$ = 199.40 Å$^3$) [56], the lattice parameter of cubic A phase for $Ni_{41}Co_9Mn_{31.5}Ga_{18.5}$ is estimated as 5.853 Å, which is in good agreement with the values of the very close composition $Ni_{41}Co_9Mn_{32}Ga_{18}$ (Supplemental Information). The volume change at the MT for $Ni_{41}Co_9Mn_{31.5}Ga_{18.5}$ is significantly less than that for $Ni_{45}Co_5Mn_{36.6}In_{13.4}$, which reaches about 2.4% [19]. The small volume change at MT for $Ni_{41}Co_9Mn_{31.5}Ga_{18.5}$ indicates the small energy barrier between A and M phase, which is consistent with small magnetic field hysteresis in the *M-H* curves obtained using the pulsed high magnetic fields [38].

It is noteworthy that positive magnetostriction is obtained not only near $H_{M \to A}$ and $H_{A \to M}$, but also in the magnetic-field-induced A phase. These results imply magnetic field variation of the electronic and lattice entropies in the magnetic-field-induced A phase. Therefore, we herein consider magnetostriction for itinerant ferromagnets to provide fundamental understanding of how the electronic and lattice degrees of freedom contribute to entropy through magnetostriction. When applying a magnetic field to an itinerant ferromagnet, the magnetization increases, leading to an increase in the Coulomb repulsive force among magnetic atoms. The distance between the magnetic atoms is determined self-consistently by the balance between the Coulomb repulsive force and the attractive force among the band electrons. When the distance between the magnetic atoms increases concomitantly with increasing applied magnetic field (i.e., the band-width decreases), the DOS increases, whereas the Debye temperature decreases [57,58]. Both electronic and lattice degrees of freedom can contribute to the inverse MCE through positive magnetostriction.

Herein, we briefly summarize the findings obtained from the magnetization, MCE, specific heat, and magnetostriction measurements. For $Ni_{41}Co_9Mn_{31.5}Ga_{18.5}$, the inverse MCEs were obtained not only at the MFIMT, but also in the magnetic-field-induced A phase. Magnetization measurements revealed that the magnetic entropy of the ferromagnetic A phase decreases concomitantly with increasing magnetic field, which contributes to direct MCE. Instead, the specific heat and magnetostriction measurements revealed that the electronic and the lattice entropy changes can contribute to inverse MCE in the magnetic-field-induced A phase through the positive magnetostriction.

**E. Theoretical calculation**

To elucidate the microscopic origins of the inverse MCE of the magnetic-field-induced A phase, we present the theoretically obtained electronic, magnetic, and structural properties. First, we obtain the lattice parameter and the ground state magnetic configurations using VASP. As described in Sec. II, we used 100 atoms unit cell of $Ni_{41}Co_9Mn_{32}Ga_{18}$ and $Ni_{45}Co_5Mn_{37}In_{13}$, which have the closest compositions to the experimentally studied samples. We calculated the total energy for the cubic $L2_1$ structure. As shown in the inset of Fig. 6(a), the $L2_1$ structure of Heusler alloy $(X_1X_2)YZ$ consists of four interpenetrating face-centered-cubic (fcc) sublattices with origin at the fractional positions: $X_1$ = (0 0 0), $X_2$ = (0.5 0.5 0.5), Y = (0.25 0.25 0.25), and Z = (0.75 0.75 0.75). For $Ni_{41}Co_9Mn_{32}Ga_{18}$ and $Ni_{45}Co_5Mn_{37}In_{13}$, the $X_1$ site is occupied by Ni atom, whereas the $X_2$ site is occupied by both Ni and Co atoms. The Y site is occupied by Mn atom. The Z site is occupied by both Mn and Ga (In) atoms. Here, $Ni_1$ and $Ni_2$ respectively denote the Ni atoms at $X_1$ and $X_2$ sites. Similarly, $Mn_1$ and $Mn_2$ denote the Mn atoms at Y and Z sites. For the ferromagnetic state, local minima in the energy profiles are obtained at $a$ = 5.808 Å for $Ni_{41}Co_9Mn_{32}Ga_{18}$ and at $a$ = 5.970 Å for $Ni_{45}Co_5Mn_{37}In_{13}$ as shown in Fig. S-1 in Supplemental Information. The calculated lattice parameters are in good agreement with the experimentally obtained ones for very similar compounds: $Ni_{41}Co_9Mn_{31}Ga_{19}$ (Supplemental Information) and $Ni_{45}Co_5Mn_{36.6}In_{13.4}$ [19].

Subsequently, we present the tetragonal distortion dependences of the total energy with respect to the ferromagnetic cubic phase ($c/a$ = 1.0) keeping the volume fixed to that of the cubic phase, as shown in Figs. 6(a) and 6(b). In both compounds, the local minima at $c/a$ = 1.00 are obtained when the system is in ferromagnetic state, which is consistent with the experimentally obtained *M-H* curves [23,38]. When the system is in a ferrimagnetic state, in which the magnetic moments of $Mn_2$ are antiparallel to those of $Mn_1$, the local minima at $c/a$ = 1.25 are obtained for both compounds. Orlandi *et al*. reported that, via the neutron diffraction data, the ground state of M phase for $Ni_{41}Co_9Mn_{32}Ga_{18}$ is antiferromagnetic, in which the magnetic moments in X, Y, and Z sublattices are imposed to order antiferromagnetically by the magnetic symmetry [59]. However, our calculations of the total energy suggest that the antiferromagnetic state is not energetically favored for Ni$_{41}$Co$_9$Mn$_{32}$Ga$_{18}$, as presented by the open triangles in Fig. 6(a). Here, we considered two types of magnetic configurations for the antiferromagnetic state: (1) the magnetic moments in Mn$_1$ and Mn$_2$ sublattices align, respectively, ferromagnetically in the *ac* plane and antiferromagnetically along the *b* axis which is the same as that proposed in Ref. [59]; (2) 50% of the magnetic moments in Mn$_1$(Mn$_2$) atoms are antiparallel to the rest in Mn$_1$(Mn$_2$) atoms, which is presented in Fig. 6(a). We find that for the M phase the latter one is energetically more favorable compared to the former one. It is noteworthy that, because the lattice constants along the *a* and *b* axes for *c*/*a* > 1 is smaller than that for *c*/*a* = 1, the ferromagnetic ordering along the *a* axis is not favored for the M phase. Thus, the latter magnetic configuration, in which the magnetic moments in Mn$_1$ and Mn$_2$ sublattices can, respectively, couple antiferromagnetically along *a*, *b*, and *c* axes, is energetically favored. Furthermore, because the antiferromagnetic coupling between Mn$_1$ and Mn$_2$ is stronger than that between Mn$_1$(Mn$_2$) and Mn$_1$(Mn$_2$), the ferrimagnetic configuration, in which all magnetic moments of Mn$_2$ are antiparallel to those of Mn$_1$, is energetically more favored. Consequently, the ferrimagnetic M phase is theoretically expected to realize at *c*/*a* = 1.25, as presented in Fig. 6(a).

The experimentally obtained magnetization for Ni$_{41}$Co$_9$Mn$_{31.5}$Ga$_{18.5}$ measured at 4.2 K supports the ferrimagnetic M phase. As presented in Fig. 10, the *M-H* curve of the M phase for Ni$_{41}$Co$_9$Mn$_{31.5}$Ga$_{18.5}$ shows ferromagnetic behavior. The magnetization reaches 49.7 Am$^2$/kg at 2 T, which corresponds to 0.52 $\mu_B$/atom. When the above described antiferromagnetic state is realized at low temperature, the net magnetization for Ni$_{41}$Co$_9$Mn$_{31.5}$Ga$_{18.5}$ is estimated as almost zero, as presented in Table 1. Therefore, the experimentally obtained spontaneous magnetization of the M phase cannot be explained by the antiferromagnetic ground state. In addition, the small temperature variation of the magnetic susceptibility indicates the absence of the paramagnetic impurities, as presented in Fig. 1(c). Therefore, for this study, we adopt the ferrimagnetic ground state of M phase for Ni$_{41}$Co$_9$Mn$_{32}$Ga$_{18}$.

As described in Sec. III A, $Ni_{41}Co_9Mn_{31.5}Ga_{18.5}$ undergoes magnetic phase transition at $T_M^C$ = 263 K. Therefore, it is expected that the ferrimagnetic state is realized below 263 K in the M phase of $Ni_{41}Co_9Mn_{31.5}Ga_{18.5}$. In contrast, the experimentally obtained *M-H* curves for $Ni_{45}Co_5Mn_{36.7}In_{13.3}$ and $Ni_{45}Co_5Mn_{36.5}In_{13.5}$ measured at 4.2 K show paramagnetic behavior. Furthermore, the temperature dependence of the specific heat for $Ni_{45}Co_5Mn_{36.5}In_{13.5}$ indicates that long range magnetic ordering does not occur in M phase, at least above 2 K [23]. Consequently, the M phase of $Ni_{45}Co_5Mn_{36.7}In_{13.3}$ and $Ni_{45}Co_5Mn_{36.5}In_{13.5}$ is paramagnetic. Although the theoretically calculated ferrimagnetic M phase for $Ni_{45}Co_5Mn_{36.7}In_{13.3}$ is not realized in real materials, we hereinafter adopt the ferrimagnetic ground state for calculations of $Ni_{45}Co_5Mn_{37}In_{13}$ to obtain more universal understanding of the electronic and magnetic properties in Ni-Mn based Heusler alloys. It is noteworthy that the total energy of the ferrimagnetic M phase at $c/a$ = 1.25 for $Ni_{41}Co_9Mn_{32}Ga_{18}$ is remarkably lower than that of ferromagnetic A phase at $c/a$ = 1, contrary to the case of $Ni_{45}Co_5Mn_{37}In_{13}$, in which the total energy of the ferrimagnetic M phase is slightly higher than that of ferromagnetic A phase. Furthermore, the energy difference at $c/a$ = 1 between a ferromagnetic and ferrimagnetic state is considerably smaller than that of $Ni_{45}Co_5Mn_{37}In_{13}$. These results indicate that the ferromagnetic A phase of $Ni_{41}Co_9Mn_{32}Ga_{18}$ is more unstable than that of $Ni_{45}Co_5Mn_{37}In_{13}$. It is also noteworthy that ferromagnetic state is favored in the $c/a$ range of 1.00-1.10, whereas the ferrimagnetic state is favored in the $c/a$ range of 1.10-1.25 for both compounds.

Tables 1 and 2 present the magnetic moments of each atom for $Ni_{41}Co_9Mn_{31.5}Ga_{18.5}$ and $Ni_{45}Co_5Mn_{36.7}In_{13.3}$ calculated using SPR-KKR. It must be described that the compositions studied in SPR-KKR and VASP are very similar. The total magnetic moments in the ferromagnetic A phase are calculated, respectively, as 1.415 $\mu_B$/atom for $Ni_{41}Co_9Mn_{31.5}Ga_{18.5}$ and 1.667 $\mu_B$/atom for $Ni_{45}Co_5Mn_{36.7}In_{13.3}$, which show good agreement with that estimated from experimentally obtained *M-H* curves: 1.33 $\mu_B$/atom for $Ni_{41}Co_9Mn_{31.5}Ga_{18.5}$ [38] and 1.68 $\mu_B$/atom for $Ni_{45}Co_5Mn_{36.7}In_{13.3}$ [23].

The total magnetic moment of $Ni_{45}Co_5Mn_{36.7}In_{13.3}$ is about 18% larger than that of $Ni_{41}Co_9Mn_{31.5}Ga_{18.5}$, mainly because of the differences of the magnetic moments of $Mn_1$ and $Mn_2$ atoms between the two compounds. As described above, because the lattice parameter of $Ni_{45}Co_5Mn_{37}In_{13}$ is larger than that of $Ni_{41}Co_9Mn_{32}Ga_{18}$, the $d$-electrons in the Mn atoms of $Ni_{45}Co_5Mn_{37}In_{13}$ are more localized than that of $Ni_{41}Co_9Mn_{32}Ga_{18}$. These localized characteristics of the $d$-electrons enlarge the total magnetic moments in the ferromagnetic A phase of $Ni_{45}Co_5Mn_{36.7}In_{13.3}$.

Next, we present the energy dependence of DOS for $Ni_{41}Co_9Mn_{31.5}Ga_{18.5}$ and for $Ni_{45}Co_5Mn_{36.7}In_{13.3}$. Here, $Ni_{1.64}Co_{0.36}Mn_{1.26}Ga_{0.74}$ and $Ni_{1.8}Co_{0.2}Mn_{1.47}In_{0.53}$ are defined as the formula units. Figures 7(a) and 7(f) respectively show the spin polarized total DOS for $Ni_{41}Co_9Mn_{31.5}Ga_{18.5}$ and for $Ni_{45}Co_5Mn_{36.7}In_{13.3}$. Here, the black solid curve represents DOS for the ferromagnetic A phase. The red dotted curve represents DOS for the ferrimagnetic M phase. Figures 7(b)-(e) and 7(g)-(j) show the decomposed DOSs for respective atoms. In the minority spin band of cubic A phase, sharp peaks are apparent near the Fermi level ($E = 0$) as indicated by the arrows in Figs. 7(b) and 7(g), which is attributed to the Ni $3d$ $e_g$ orbital. When the system undergoes MT, this peak splits through the tetragonal distortion. Consequently, the DOS at the Fermi level in the minority spin band decreases remarkably, contrary to the majority spin bands, in which the DOSs at the Fermi level are insensitive to the tetragonal distortion. The differences of the total DOS between A and M phases is 0.782 states/eV/f.u. for $Ni_{41}Co_9Mn_{31.5}Ga_{18.5}$ and is 1.044 states/eV/f.u. for $Ni_{45}Co_5Mn_{36.7}In_{13.3}$. These values respectively reach about 42% and 49% of the total DOS of A phases. As described in Sec. III C, the specific heat measurements for $Ni_{40.3}Co_{9.7}Mn_{31.5}Ga_{18.5}$ revealed that $\gamma_M$ is about 39% smaller than $\gamma_A$, which is in good agreement with the calculated one. However, for $Ni_{45}Co_5Mn_{36.5}In_{13.5}$, $\gamma_A - \gamma_M$ is only 9% of $\gamma_A$ [23]. For $Ni_{45}Co_5Mn_{36.5}In_{13.5}$, this deviation between the experiment and the theoretical calculation is related to the differences of the magnetic structure in M phase between them.

To obtain a better understanding of the tetragonal distortion dependence of the DOS, we present the total DOS's calculated at $c/a$ = 1.00, 1.05, 1.10, 1.15, 1.20, and 1.25 in Figs. 8(a)-8(d). Here, we adopt the ferromagnetic state for the calculations of $c/a$ = 1.00, 1.05, 1.10, 1.15, and 1.20, and the ferrimagnetic state for the calculation of $c/a$ = 1.25. For both compounds, the DOS decreases concomitantly with increasing $c/a$ up to 1.20. Then, it increases abruptly at $c/a$ = 1.25. To understand this behavior, we consider the contribution of the microscopic magnetic structure to the DOS in M phase. When the system is in a ferrimagnetic state, the spin of $Mn_2$ is aligned antiparallel to the other spins. In such a case, the 3$d$ electrons of Ni atoms become more itinerant through the superexchange path of Ni-$Mn_2$-Ni, which contributes to increase of the DOS. Consequently, the DOS becomes insensitive to a change in $c/a$ in the range of 1.10 to 1.25. For the same reason, it is expected that, when the system is in paramagnetic state, the DOS becomes more insensitive to the tetragonal distortion. Therefore, the earlier reported small change of $\gamma$ in $Ni_{45}Co_5Mn_{36.5}In_{13.5}$ can be attributed to the paramagnetic M phase [23].

## IV. DISCUSSION

Next, we discuss the electronic origin of the inverse MCE in the magnetic-field-induced A phase of $Ni_{41}Co_9Mn_{31.5}Ga_{18.5}$. Figure 9 shows the change in the total DOS at the Fermi energy as a function of $c/a$ [$\Delta D_{FE}(c/a)$] for $Ni_{41}Co_9Mn_{31.5}Ga_{18.5}$. Because the ferrimagnetic state is favored in the range of 1.10-1.25, as described above, the values at 1.15 and 1.20 calculated for ferromagnetic state are incorrect. For that reason, we ignore these data, and expect a gradual change in DOS in the range of 1.10-1.25 as indicated by the dotted curve in Fig. 9. It is noteworthy that about 78% of the total change in DOS occurs in the range of 1.00-1.10. If applying a uniaxial stress along the $c$ axis to the cubic A phase, and controlling the $c/a$ in the range of 1.00-1.10, then the system changes reversibly because $c/a = 1.10$ is a critical point of the elastic-plastic displacement. In this case, the elastocaloric effect attributable to the electronic entropy change is expected to be obtained.

By contrast, in the magnetic-field-induced A phase, the crystal structure is self-consistently determined by the balance among the exchange energy between the spins, the elastic energy of the crystal, and the Zeeman energy, as described in Sec. III D. As depicted in Fig. 6(a), for $Ni_{41}Co_9Mn_{31.5}Ga_{18.5}$, the total energy of the ferromagnetic state is close to that of the ferrimagnetic state in the $c/a$ range of 1.00-1.10. Therefore, it is expected that the tetragonal distortion can be induced easily by the disorder of spins in the $Mn_1$ and $Mn_2$ sites in the magnetic-field-induced A phase. Figure 10 portrays the $M$-$H$ curve of $Ni_{41}Co_9Mn_{31.5}Ga_{18.5}$ measured at 4.2 K [38]. It is noteworthy that the magnetization is not saturated in magnetic fields up to 56 T. In fact, it gradually increases concomitantly with increasing magnetic field with the slope of $dM/\mu_0 dH \sim 0.1$ Am$^2$/kg/T, which is about five times larger than that of $Ni_{45}Co_5Mn_{36.7}In_{13.3}$ (i.e., the disorder of spins partially remains) [23]. This large slope of magnetization also demonstrates the large DOS change at the Fermi level of $Ni_{41}Co_9Mn_{31.5}Ga_{18.5}$, which is the characteristic of itinerant ferromagnetic Heusler alloys [60]. The magnetization at 56 T is about 130 Am$^2$/kg (1.372 $\mu_B$/atom), which is still smaller than the theoretically obtained total magnetic moment of the ferromagnetic A phase. As described in Sec. III E, for $Ni_{45}Co_5Mn_{36.7}In_{13.3}$, the experimentally obtained magnetization in the magnetic-field-induced A phase reaches the theoretically obtained total magnetic moment in the ferromagnetic A phase. This indicates that the partial disorder of spins does not occur in $Ni_{45}Co_5Mn_{36.7}In_{13.3}$, which is consistent with the large energy difference between ferromagnetic and ferrimagnetic states in the $c/a$ range of 1.00-1.10 [Fig. 6(a)].

As described above, the crystal structure of the magnetic-field-induced A phase for $Ni_{41}Co_9Mn_{31.5}Ga_{18.5}$ is coupled strongly to its magnetic structure. Contrastingly, the ferromagnetic cubic A phase is stable for $Ni_{45}Co_5Mn_{36.7}In_{13.3}$. For $Ni_{41}Co_9Mn_{31.5}Ga_{18.5}$, it is expected that the reversible magnetostriction in the *c/a* range of 1.00-1.10 is induced by the magnetic field, as depicted schematically in the insets of Fig. 10. Therefore, the inverse MCEs observed in the magnetic-field-induced A phase for $Ni_{41}Co_9Mn_{31.5}Ga_{18.5}$ are explainable by the electronic entropy change attributable to the magnetic field variation of the DOS through reversible magnetostriction. Such reversible magnetostrictions are indeed apparent in Fig. 5. However, discussing the absolute value of the change in *c/a* through the present data is difficult because the data presented in Fig. 5 were taken using a polycrystalline sample. For that reason, the clear evidence of such reversible structural change remains insufficient.

## V. CONCLUDION

The inverse MCE of $Ni_{41}Co_9Mn_{31.5}Ga_{18.5}$ was measured directly using pulsed high magnetic fields up to 56 T. This effect is observed not only for MFIMTs, but also for the magnetic-field-induced A phase, which shows a reversible adiabatic temperature change. The specific heat and magnetostriction measurements revealed that the electronic and the lattice entropy changes contribute to the inverse MCE in the magnetic-field-induced A phase. Through theoretical analysis using the first-principles calculations of the total energy and the DOSs, we found the following: (1) for $Ni_{41}Co_9Mn_{31.5}Ga_{18.5}$, the energy barrier between A and M phase is quite small; (2) the A phase of $Ni_{41}Co_9Mn_{31.5}Ga_{18.5}$ is more itinerant than that of $Ni_{45}Co_5Mn_{36.7}In_{13.3}$; (3) because, in the *c/a* range of 1.00-1.10, the total energy of the ferromagnetic state is close to that of the ferrimagnetic state, the magnetic-field-induced ferromagnetic A phase is unstable. For the reasons described above, the tetragonal distortion can be induced easily by the partial retention of the disorder of spins when the externally applied magnetic field is insufficient. Consequently, the DOS of the Ni 3*d* $e_g$ minority spin states can be controllable by the externally applied magnetic field through the reversible magnetostriction in the magnetic-field-induced A phase. Therefore, we conclude that the electronic entropy change attributable to the magnetic field variation of the DOS plays a dominant role in the inverse MCE of the magnetic-field-induced A phase of $Ni_{41}Co_9Mn_{31.5}Ga_{18.5}$. To obtain clear evidence of the reversible change in the crystal structure of the magnetic-field-induced A phase, microscopic experiments such as X-ray diffraction and neutron diffraction in high magnetic fields are desired. Furthermore, detailed investigation of the lattice entropy change, which can also contribute to the inverse MCE through the tetragonal distortion, is important for elucidating the unconventional magnetocaloric properties of this material.

**ACKNOWLEDGMENTS**

This work was partly supported by the Ministry of Education, Culture, Sports, Science, and Technology, Japan, through a Grant-in-Aid for Early Career Scientist (Grant No. 18K13979) and the Fuji Science and Technology Foundation. Some of this work was performed through the joint research with the Institute for Solid State Physics, The University of Tokyo and with the High Field Laboratory for Superconducting Materials, Institute for Materials Research, Tohoku University.

Table 1. Magnetic moment of each ion in the A and M phase for $Ni_{41}Co_9Mn_{31.5}Ga_{18.5}$.

|  | $Ni_1$ ($\mu_B$) | $Ni_2$ ($\mu_B$) | Co ($\mu_B$) | $Mn_1$ ($\mu_B$) | $Mn_2$ ($\mu_B$) | Ga ($\mu_B$) | Total ($\mu_B$/atom) |
|---|---|---|---|---|---|---|---|
| Cubic (FM) | 0.540 | 0.507 | 1.263 | 3.476 | 3.567 | -0.081 | 1.415 |
| Tetragonal (FIM) | 0.259 | 0.272 | 0.717 | 3.363 | -3.547 | -0.074 | 0.769 |
| Tetragonal (AFM) | 0.002 | 0.002 | 0.002 | 3.451 / -3.481 | 3.498 / -3.528 | 0 | -0.004 |

Table 2. Magnetic moment of each ion in the A and M phase for $Ni_{45}Co_5Mn_{36.7}In_{13.3}$.

|  | $Ni_1$ ($\mu_B$) | $Ni_2$ ($\mu_B$) | Co ($\mu_B$) | $Mn_1$ ($\mu_B$) | $Mn_2$ ($\mu_B$) | In ($\mu_B$) | Total ($\mu_B$/atom) |
|---|---|---|---|---|---|---|---|
| Cubic (FM) | 0.559 | 0.552 | 1.336 | 3.654 | 3.810 | -0.079 | 1.667 |
| Tetragonal (FIM) | 0.190 | 0.197 | 0.739 | 3.581 | -3.879 | -0.057 | 0.556 |

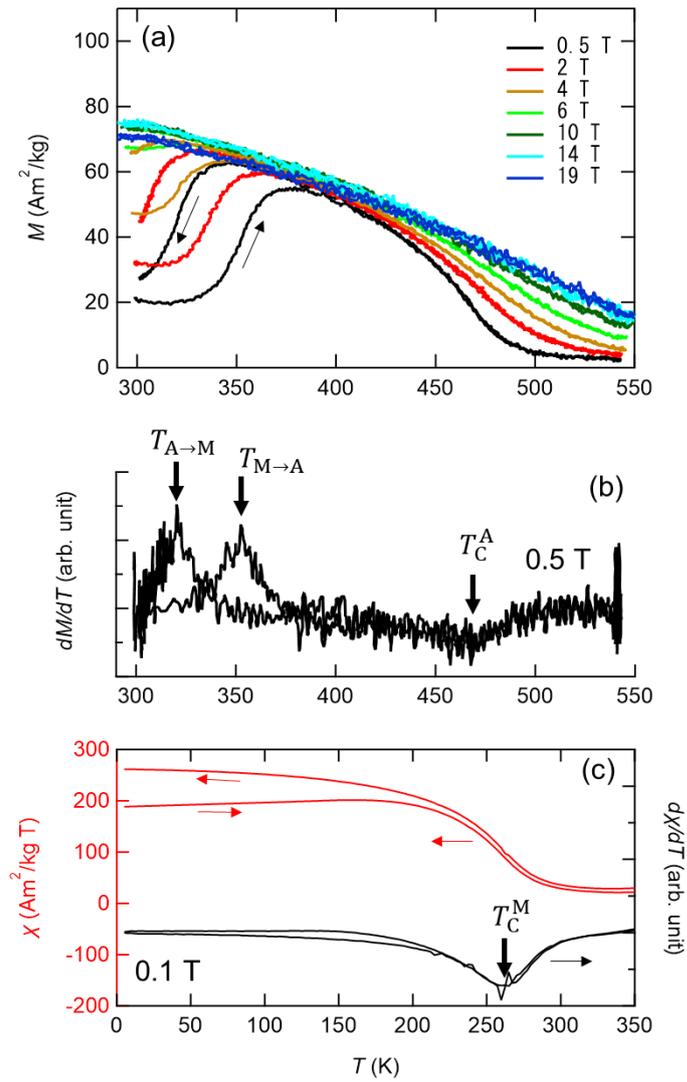

Fig. 1. (a) *M-T* curves of $Ni_{41}Co_9Mn_{31.5}Ga_{18.5}$ measured at various magnetic fields. (b) Temperature differential of the *M-T* curve at 0.5 T. (c) Temperature dependence of magnetic susceptibility for $Ni_{41}Co_9Mn_{31.5}Ga_{18.5}$ measured at 0.1 T and its temperature differential.

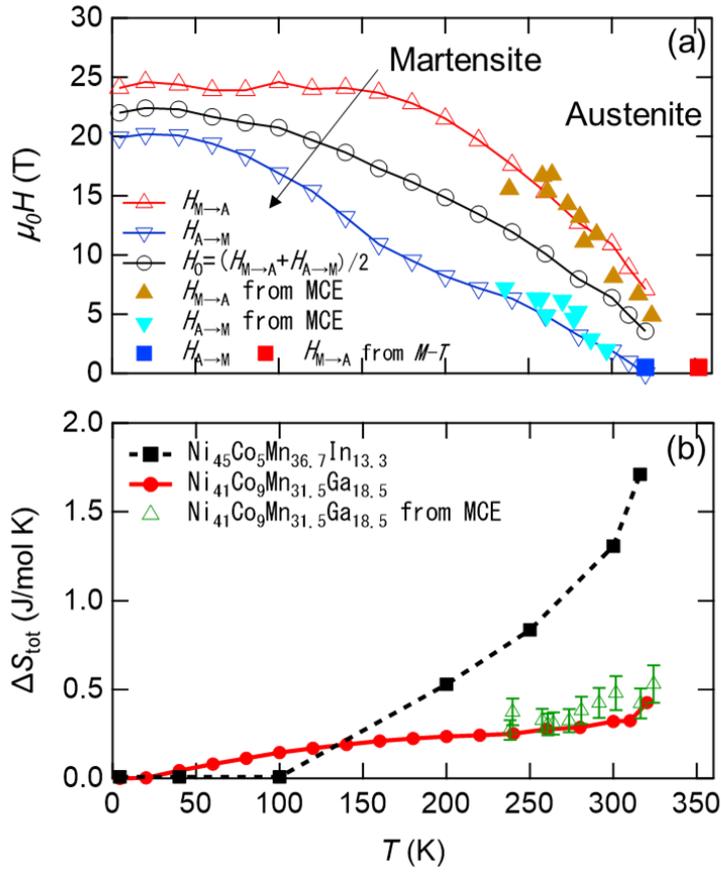

Fig. 2. (a) Magnetic phase diagram of $Ni_{41}Co_9Mn_{31.5}Ga_{18.5}$ determined by magnetization and MCE measurements. (b) Temperature dependence of the total entropy change for $Ni_{41}Co_9Mn_{31.5}Ga_{18.5}$ and for $Ni_{45}Co_5Mn_{36.7}In_{13.3}$ [23] estimated, respectively, from the magnetic phase diagrams and from the $\Delta T_{MT}$ presented in Fig. 3(c).

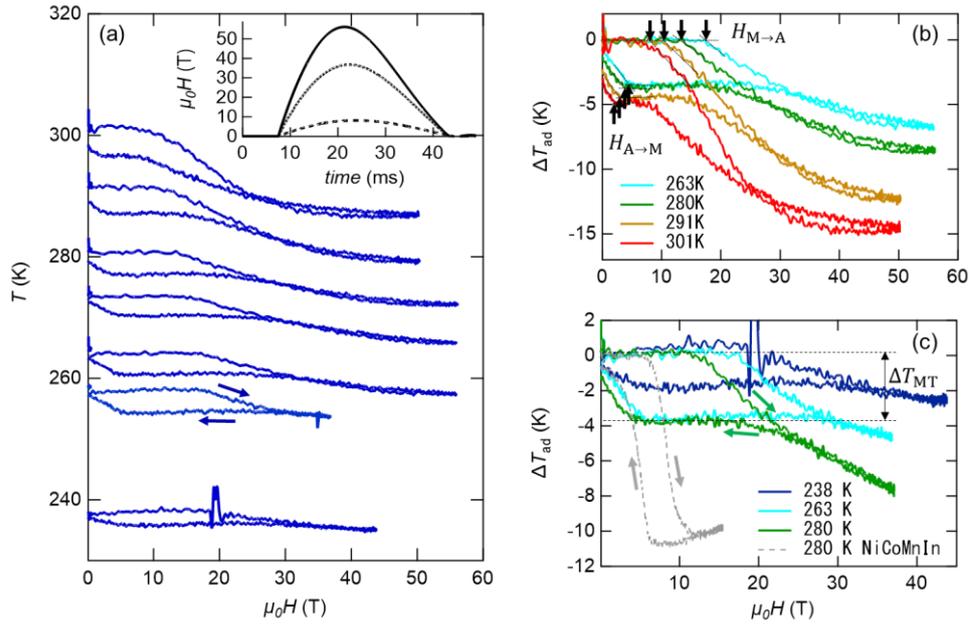

Fig. 3. (a) Adiabatic temperature changes of $Ni_{41}Co_9Mn_{31.5}Ga_{18.5}$ polycrystalline sample as functions of the applied magnetic field. The magnetic field profiles are shown in the inset. Also, (b) and (c) present enlarged views of the representative temperatures, in which the $T_{ini}$ is subtracted from each datum. The gray dotted curve shows a result of $Ni_{45}Co_5Mn_{36.7}In_{13.3}$ measured at $T_{ini}$ = 280 K [23].

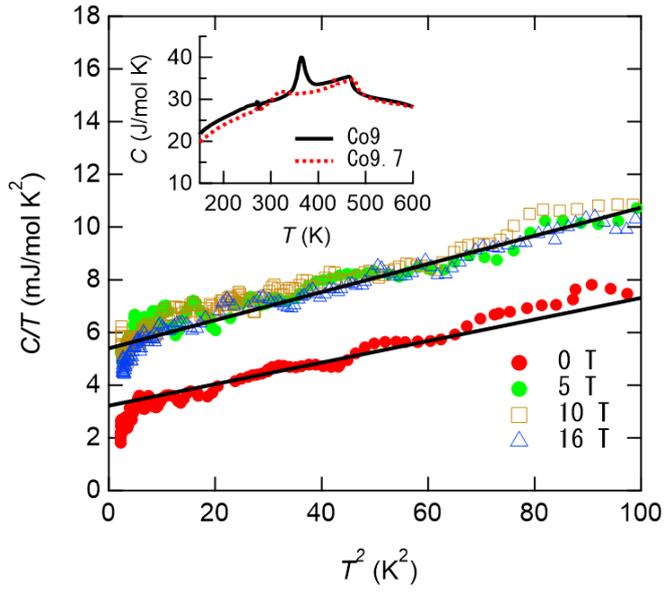

Fig. 4. Specific heat of $Ni_{40.3}Co_{9.7}Mn_{31.5}Ga_{18.5}$ shown as $C/T$ vs. $T^2$. Data measured at 5, 10, and 16 T were taken after field cooling at 17.5 T (i.e., the sample is in A phase). The inset presents results obtained at around $T_{M \to A}$ for $Ni_{41}Co_9Mn_{31.5}Ga_{18.5}$ (Co9) and $Ni_{40.3}Co_{9.7}Mn_{31.5}Ga_{18.5}$ (Co9.7) taken using a DSC shown as $C$ vs. $T$.

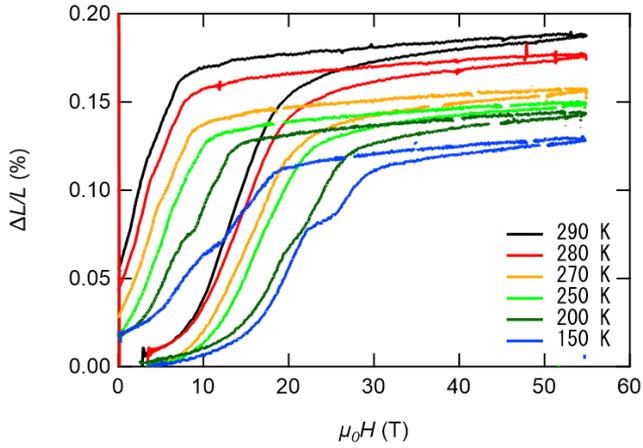

Fig. 5. Magnetostrictions of $Ni_{41}Co_9Mn_{31.5}Ga_{18.5}$ polycrystalline sample measured along the applied magnetic field ($\mu_0 H \parallel \Delta L$).

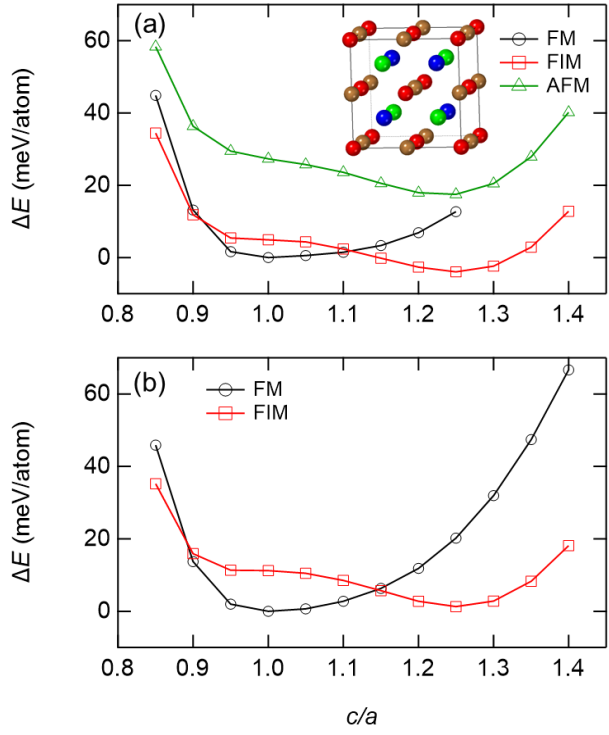

Fig. 6 Total energy shown as a function of the ratio of the tetragonality ($c/a$) for (a) $Ni_{41}Co_9Mn_{32}Ga_{18}$ and for (b) $Ni_{45}Co_5Mn_{37}In_{13}$. FM, FIM, and AFM respectively denote the ferromagnetic, ferrimagnetic, and antiferromagnetic. The origin of the energy scale is set to the energy at $c/a = 1$ for FM. The inset of (a) portrays the crystal structure of cubic ($L2_1$) A phase in the Heusler alloy ($X_1X_2$)YZ ($X_1$, blue; $X_2$, green; Y, red; Z, brown).

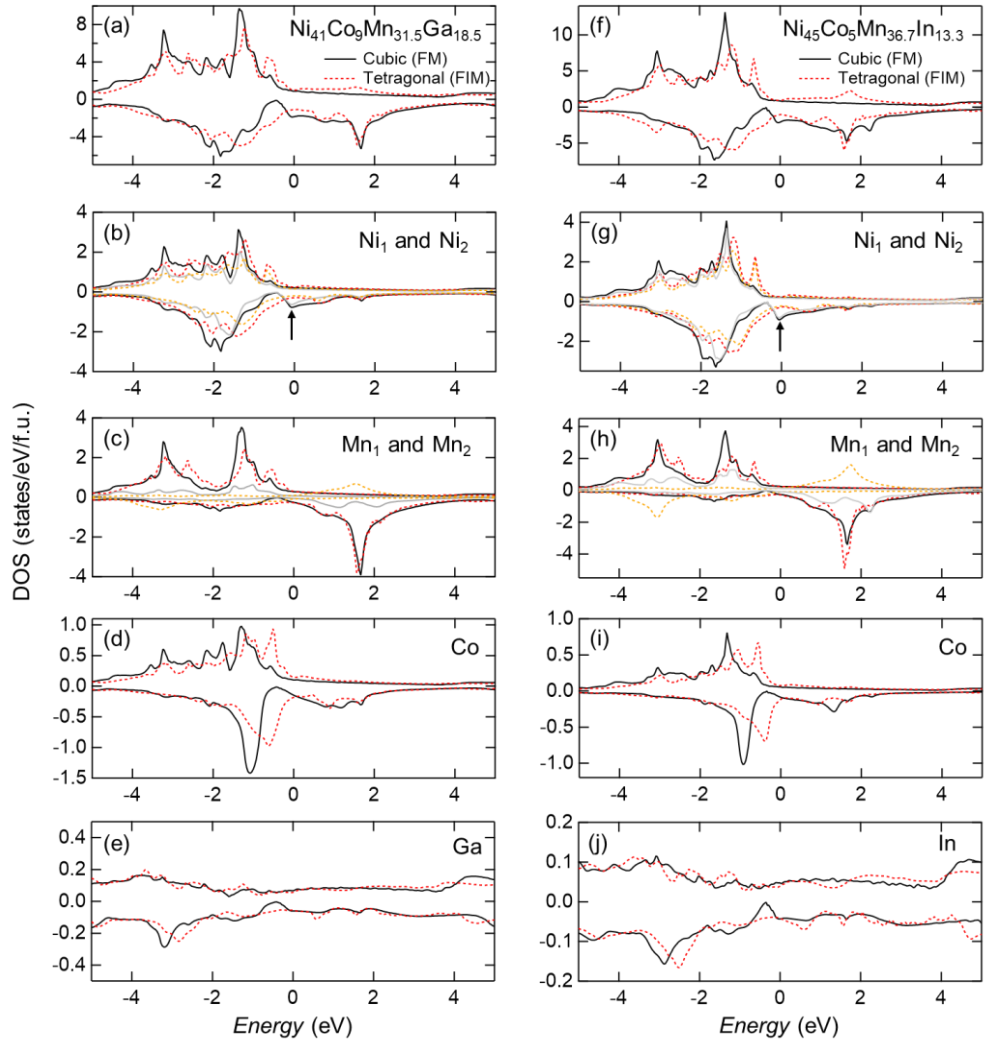

Fig. 7 Density of states calculated for $Ni_{41}Co_9Mn_{31.5}Ga_{18.5}$ [(a)-(e)] and for $Ni_{45}Co_5Mn_{36.7}In_{13.3}$ [(f)-(j)]. Solid and dotted curves represents, respectively, cubic (ferromagnetic, FM) and tetragonal (ferrimagnetic, FIM) phases. Gray solid and brown dotted curves are the DOSs of $Ni_2$ and $Mn_2$.

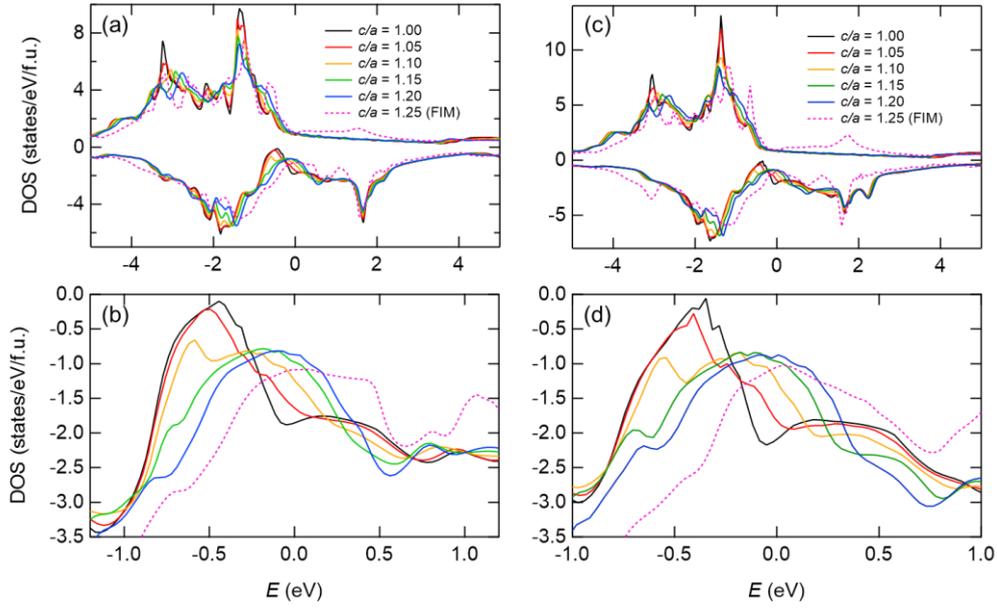

Fig. 8 *c/a* dependence of the DOS for $Ni_{41}Co_9Mn_{31.5}Ga_{18.5}$ [(a) and (b)] and for $Ni_{45}Co_5Mn_{36.7}In_{13.3}$ [(c) and (d)]. Also, (b) and (d) present the enlarged views of the minority spin band at around the Fermi level.

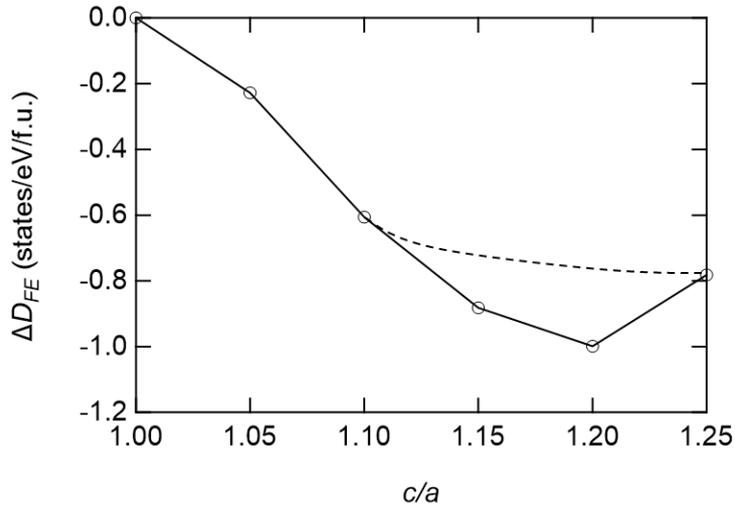

Fig. 9 *c/a* dependence of the change in DOS [$\Delta D_{FE} = D_{FE}(1.00) - D_{FE}(c/a)$] for $Ni_{41}Co_9Mn_{31.5}Ga_{18.5}$.

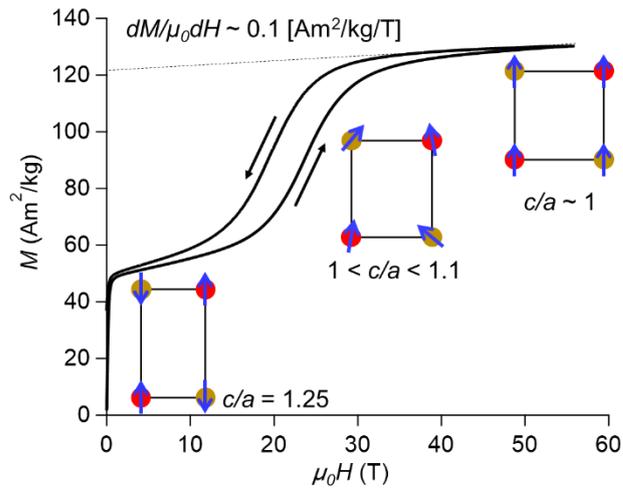

Fig. 10 *M-H* curve of $Ni_{41}Co_9Mn_{31.5}Ga_{18.5}$ measured at 4.2 K [38]. The insets are projection of the crystal structure along the *a* axis, where the red and brown circles respectively denote $Mn_1$ and $Mn_2$ atoms.

# Supplemental information for first principles calculation of the electronic, magnetic, and structural properties for $Ni_{41}Co_9Mn_{32}Ga_{18}$ and $Ni_{45}Co_5Mn_{37}In_{13}$ alloys

T. Kihara[1]*, T. Roy[2], X. Xu[3], A. Miyake[4], M. Tsujikawa[2], H. Mitamura[4] M. Tokunaga[4], Y. Adachi[5], T. Eto[6], and T. Kanomata[7]

[1] Institute for Materials Research, Tohoku University, Sendai, Miyagi, Japan
[2] Research Institute of Electrical Communication, Tohoku University, Sendai, Miyagi, Japan
[3] Department of Materials Science, Tohoku University, Sendai, Miyagi, Japan
[4] The Institute for Solid State Physics, The University of Tokyo, Kashiwa, Chiba, Japan
[5] Graduate School of Science and Engineering, Yamagata University, Yonezawa, Yamagata, Japan
[6] Kurume Institute of Technology, Kurume, Fukuoka, 830-0052, Japan
[7] Research Institute for Engineering and Technology, Tohoku Gakuin University, Tagajo, Miyagi, Japan

*Address all correspondence to: t_kihara@imr.tohoku.ac.jp

As Supplemental Information, we compare the lattice parameter variation of the total energy and the result of powder X-ray diffraction (PXRD) for $Ni_{41}Co_9Mn_{32}Ga_{18}$ to demonstrate consistency of the calculations. Fig. S-1 presents the total energy shown as a function of the cubic unit cell volume. Calculations were performed for the $L2_1$ structures of $Ni_{41}Co_9Mn_{32}Ga_{18}$ and of $Ni_{45}Co_5Mn_{37}In_{13}$. Both compounds respectively portray the local minima at $V$ = 195.92 Å ($a$ = 5.808 Å) for $Ni_{41}Co_9Mn_{32}Ga_{18}$ and at $V$ = 212.78 Å ($a$ = 5.970 Å) for $Ni_{45}Co_5Mn_{37}In_{13}$.

Fig. S-2 presents the PXRD pattern obtained for $Ni_{41}Co_9Mn_{32}Ga_{18}$ alloy. All peaks observed in the angular range of $2\theta$ from 10° to 45° are well assigned to reflections from $L2_1$ austenite phase. The lattice parameter is estimated as $a$ = 5.829 Å, which shows good agreement with the theoretically obtained one for $Ni_{41}Co_9Mn_{32}Ga_{18}$. However, the lattice parameter calculated for $Ni_{45}Co_5Mn_{37}In_{13}$ shows good agreement with the experimentally obtained lattice parameter of $Ni_{45}Co_5Mn_{36.6}In_{13.4}$ ($a$ = 5.978 Å), as reported by Kainuma *et al* in Ref. [2].

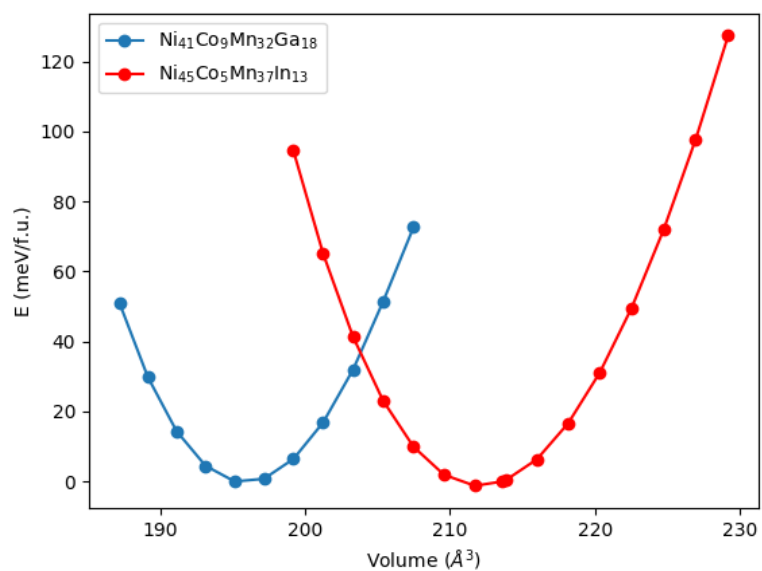

Fig. S-1 Total energy of cubic $L2_1$ phase for $Ni_{41}Co_9Mn_{32}Ga_{18}$ and for $Ni_{45}Co_5Mn_{37}In_{13}$ shown as a function of the volume of a cubic unit cell.

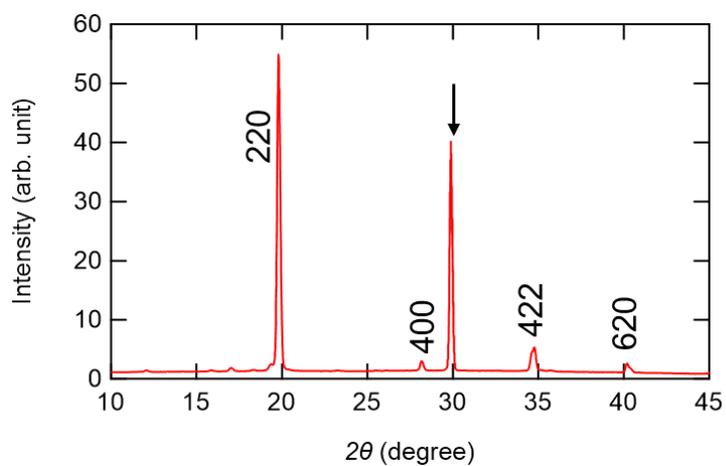

Fig. S-2 PXRD pattern of $Ni_{41}Co_9Mn_{31}Ga_{19}$ measured at room temperature. The peak indicated by the arrow is attributed to the (110) reflection from the Re gasket.